\documentclass[aps,pra,twocolumn,groupedaddress,showpacs]{revtex4-1}
\usepackage{color,graphicx}
\usepackage{dcolumn}
\usepackage{bm}
\usepackage{gensymb}
\usepackage{amsmath}
\usepackage{siunitx}

\begin{document}

\title{Production of large Bose-­Einstein condensates in a magnetic-shield-compatible hybrid trap}
\author{Giacomo Colzi$^{1,2}$}
\email[]{giacomo.colzi@unitn.it}
\author{Eleonora Fava$^{1}$}
\author{Matteo Barbiero$^{1,3,4}$}
\author{Carmelo Mordini$^{1,2}$}
\author{Giacomo Lamporesi$^{1,2}$}
\author{Gabriele Ferrari$^{1,2}$}

\affiliation{$^1$ INO-CNR BEC Center and Dipartimento di Fisica, Universit\`a di Trento, 38123 Trento, Italy}
\affiliation{$^2$ Trento Institute for Fundamental Physics and Applications, INFN, 38123 Trento, Italy}
\affiliation{$^3$ Politecnico di Torino, Corso Duca degli Abruzzi, 24 10129 Torino, Italy}
\affiliation{$^4$ Istituto Nazionale di Ricerca Metrologica, Strada delle Cacce 91, Torino 10135, Italy}

\date{\today}

\begin{abstract}

We describe the production of large ${}^{23} \mathrm{Na}$ Bose-Einstein condensates in a hybrid trap characterized by a weak magnetic field quadrupole and a tightly focused infrared beam. The use of small magnetic field gradients makes the trap compatible with the state-of-the-art magnetic shields. By taking advantage of the deep cooling and high efficiency of gray molasses to improve the initial trap loading conditions, we produce condensates composed of as much as $7$ million atoms in less than  $30 \; \mathrm{s}$.

\end{abstract}
\pacs{}
%\keywords{}

\maketitle

\section {Introduction}

Ultracold atomic gas experiments emerged in recent years as a promising platform to simulate the dynamics of many-body quantum systems, thanks to the high degree of experimental control and accessibility they offer \cite{Bloch12,Georgescu14,Gross17}. Recent theoretical proposals show the possibility of using coherently coupled Bose-condensed mixtures to produce exotic defect structures exhibiting analogies with fundamental interaction physics \cite{Son02,Kasamatsu04,Kasamatsu05,Cipriani13,Tylutki16,Aftalion16,Qu17,Calderaro17,Eto17}, to study spin-orbit coupled systems \cite{Juzeliunas10,Brandon10,Li12,Li12b,Li13,Han15} allowing the realization of supersolid phases of matter \cite{Li17,Leonard17}, and to devise analog models of gravitational physics \cite{Butera17}.  

Magnetic field fluctuations and inhomogeneities are often the main sources of dephasing when coherent manipulation of internal atomic states is attempted. Employing ultracold mixtures insensitive to first order Zeeman perturbations was demonstrated to be a viable strategy to increase coherence times, for instance with $^{87}\mathrm{Rb}$ \cite{Harber02,Treutlein04,Deutsch10,Kleine11,Muessel15}, even in the absence of magnetic field screening techniques. Such mixtures, however, are characterized by interaction properties unsuitable to realize the aforementioned proposals. Conversely, the mixture composed of the ground hyperfine states $\left| 1,\pm1 \right \rangle$ of $^{23}\mathrm{Na}$ allows to devise a stable system of two BECs, perfectly overlapped in trap and characterized by a clear separation between density and spin dynamics \cite{Bienaime16,Fava17}, thanks to its favorable interaction properties, but requires to work in conditions of high magnetic field stability to preserve the internal states coherence for a sufficiently long time to study its dynamics. Similar requirements characterize many different physical systems: a few examples include experiments in electron microscopy \cite{Krivanek08}, nuclear magnetic resonance \cite{Mansfield87}, ultracold atoms \cite{Ottl06,Dedman07,Zhang15}, atomic magnetometry \cite{Sheng13} and atom interferometry \cite{VanZoest10,Milke14,Kubelka-Lange16,deAngelis09,Lamporesi08,Dickerson13,Hartwig15}.

Spurious magnetic fields can be either actively compensated or passively shielded. Active compensation is more suitable when dealing with noise in the frequency range between $10$ and $1000 \; \mathrm{Hz}$, but more challenging to implement. A good active compensation requires to monitor the field at the atoms location inside the vacuum chamber, which cannot be directly accessed, necessitating noninvasive reconstruction techniques \cite{Botti06}. Compensation of static and low-frequency external fields can be achieved using passive magnetic shields. Magnetically soft materials suitable for this application, such as $\SIUnitSymbolMicro$-metal and analogous alloys, are limited by their tendency to saturate and lose their magnetic shielding properties until a demagnetization procedure is applied. Since the elements that produce magnetic fields necessary for the experiment must be fitted inside the shield, possible incompatibility between the shield saturation limits and these elements must be taken into account. A detailed study of this subject cannot leave the specific shield and coils design out of consideration and requires a simulative approach and a performance test of the final assembly that are beyond the scope of this manuscript, and will be presented elsewhere \cite{shieldprep}.
Saturation occurs at magnetic field intensities $ |  \vec H |$ of the order of $10^2$ to $10^3 \; \mathrm{A/m}$ inside the material, depending on the specific alloy. Compressed magnetic traps exerting gradients of hundreds $\mathrm{G/cm}$ on the atoms can cause saturation issues \cite{Esslinger98,Bloch99,blochcom}, if we exclude atom chips. On the other hand, the smaller gradients that are typically used for magneto-optical trap (MOT) operation demonstrated full compatibility with properly designed shields \cite{Kim13,Sycz18}, but are insufficient to provide the necessary elastic collisions rate to reach quantum degeneracy in magnetic traps via RF evaporative cooling.

These limits can be circumvented by all-optical production protocols. Despite their well-known advantages, such as the possibility to devise spin-insensitive traps, pure optical dipole traps (ODTs) are limited by the available power in the tradeoff between the capture volume and the trap depth and, in the simplest implementations, by the reduction of trapping frequencies during evaporation.

Production of BECs in traps that combine the advantages of both optical and magnetic potentials was demonstrated to be a viable strategy since the first realizations of BEC, where a repulsive "optical plug" was used to suppress Majorana spin-flips in a simple quadrupole magnetic trap (QMT) \cite{Davis95,Naik05,Heo11,Dubessy12}. Another recently demonstrated approach consists in combining a tightly focused red-detuned single beam \cite{Lin09,Flores15,Mishra15} or a crossed \cite{Bouton15} dipole trap with a QMT. For the reasons discussed above, the adiabatic compression of the QMT remains a fundamental strategy with this second approach in order to improve the ODT loading conditions, by applying an RF evaporation stage to the magnetically trapped sample. The combined magnetic and optical potential also shows enhanced confinement along the ODT trap axis, compared to the pure optical counterpart, allowing to reach quantum degeneracy in single beam configurations even in the absence of magnetic compression  \cite{Zaiser11}. Weak magnetic field gradients can also be used to enhance all-optical production techniques, for instance to spin-polarize the sample \cite{Salomon13} or to compensate for gravity during optical evaporation \cite{Kinoshita05}.

In this work we describe the production of ${}^{23} \mathrm{Na}$ BECs in a hybrid trap composed of a low magnetic field gradient QMT and a single beam ODT, where the compression of the QMT is avoided for magnetic shield compatibility. Compared to similar realizations \cite{Zaiser11}, here we also take advantage of gray molasses (GM) cooling to efficiently load the atoms into the QMT, where they act as a reservoir during the loading of the deep (compared to the sample temperature) ODT. With such a protocol we produce condensates composed of as much as $7$ million atoms in a spin-polarized state, ensuring well controlled starting conditions for the production of internal atomic state mixtures.

\section{\label{sec:Expapp}Atomic sample preparation}

\begin{figure*}[t!]
	\resizebox{0.75\textwidth}{!}{
		\includegraphics{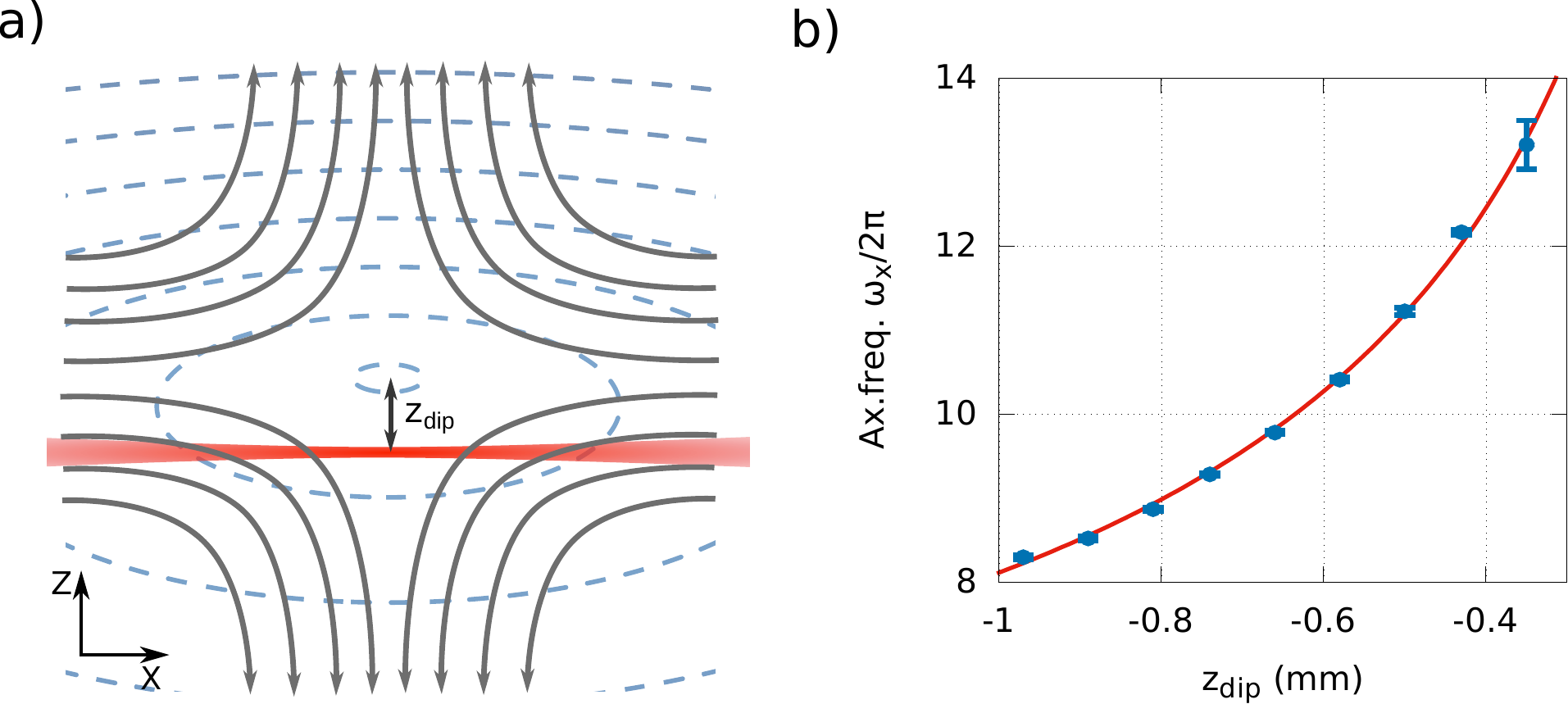}}
	\caption{\label{fig:hybridconf} \textbf{a)} Sketch of the hybrid trap configuration. A single beam dipole trap (red shaded region) is focused  at a distance $z_\mathrm{dip}$ below the quadrupole trap center. Here $z$ is the simmetry (strong) axis of the QMT, while $x$ designates the axial coordinate of the dipole beam. Magnetic field lines are pictorially represented (solid, gray) as well as magnetic potential contour lines accounting for gravity (dashed, blue). \textbf{b)} Axial trapping frequency measured as a function of the vertical displacement. The red curve is a fit to the experimental data.}
\end{figure*}

Our experimental apparatus as well as the full characterization of GM cooling were already described elsewhere \cite{Lamporesi13,Colzi16}. The atomic source is composed of a crucible where the sodium sample is evaporated, by a $12\text{-}\mathrm{cm}$-long Zeeman slower (ZS), and by a 2D MOT. The atoms loaded into the 2D MOT are pushed along its free axis by a dedicated laser beam towards the science chamber, where they are captured into a Dark-Spot (DS) MOT and GM cooling is performed.

The six orthogonal 3D MOT beams of approximate diameter of $1.9 \; \mathrm{cm}$ counterpropagate in pairs with opposite $\sigma^\pm$ polarizations and slightly red-detuned frequency from the cooling transition  ($3{}^2S_{1/2}  \left | F=2 \right \rangle \rightarrow 3{}^2P_{3/2} \left | F'=3 \right \rangle$ at $589\;\mathrm{nm}$). Atoms are repumped from the $ |F=1 \rangle $ ground state manifold by means of an additional hollow repumper beam superimposed on one of the cooling beams \cite{Ketterle93}. Such a beam is obtained by shining a collimated Gaussian beam on an axicon lens (Thorlabs AX252-A) that exchanges outer and inner parts of the beam. At a distance of $800 \; \mathrm{mm}$ from the axicon, the beam profile shows a dark disk ($6 \; \mathrm{mm}$ diameter) and a bright annular band with intensity increasing with the distance from the center. Such a profile is then imaged on the atoms  after further blocking the spurious light present in the dark region with a disk-shaped obstacle. During MOT loading, the quadrupole coils produce a gradient of $13.3 \; \mathrm{G/cm}$ at the trap center. Atoms are captured in the DS MOT at an approximate rate of  $0.6 \times 10^9 \; \mathrm{atoms/s}$, and their number saturates after a loading time of $13 \; \mathrm{s}$ to $4 \times 10^9$ at a temperature of the order of $300 \; \mathrm{\SIUnitSymbolMicro K}$. The MOT is then switched off before applying a GM cooling procedure to the atomic cloud \cite{Colzi16}.

GM cooling operates on the blue side of the D1 optical transition ($3{}^2S_{1/2}  \left | F=2 \right \rangle \rightarrow 3{}^2P_{1/2} \left | F'=2 \right \rangle$) with an additional repumper sideband amounting to roughly $4\%$ of the total power on the ($ 3{}^2S_{1/2} \left | F=1 \right \rangle \rightarrow 3{}^2P_{1/2} \left | F'=2 \right \rangle$). The six $3\text{-}\mathrm{mm}$-waist ($1/e^2$ radius) beams propagate along the same axes and with the same polarizations as the MOT beams. After a $0.5 \; \mathrm{ms}$ capture pulse detuned by $4 \Gamma$ ($\Gamma \sim 2 \pi \, \mathsf{x} \, 10 \; \mathrm{MHz} $) from the cooling transition with an intensity of $280 \; \mathrm{mW/cm^2}$ per beam, the power is ramped down  to $17 \; \mathrm{mW/cm^2} $ in $4.5 \; \mathrm{ms}$ after increasing the detuning to $12 \Gamma$. In order to efficiently depump the atom population from the $\left | F=2 \right \rangle$ manifold, the beams are operated at $4 \Gamma$ detuning and without repumper sidebands during the last $0.5 \; \mathrm{ms}$ of the ramp. Temperatures of $14 \; \SIUnitSymbolMicro \mathrm{K} $ with roughly unitary capture efficiency are obtained.

\section{\label{sec:hybridtrap}Hybrid trap}

	\begin{table*}[t]
		\begin{center}
			\caption{\label{tab:stages} Atom number, temperature, peak density and phase-space density at various stages of the experimental sequence. Reported uncertainties account for statistical errors.}
			\begin{ruledtabular}
				\begin{tabular}{lllll} 
					$\space$ & $N$ (atoms) & $ T(\SIUnitSymbolMicro \mathrm{K}) $ & $\rho_\mathrm{0} (\mathrm{atoms}/cm^{-3})$ & $PSD$  \\ 
					\hline\noalign{\smallskip}
					DS MOT & $4.0(5) \times 10^{9} $  & $310(20)$ & $1.5(4)  \times 10^{11} $ & $1.1(3) \times 10^{-6} $ \\
					GM & $4.14(2)  \times 10^{9} $ & $14.8(3)$ & $1.50(7) \times 10^{11} $  & $1.04(6)\times 10^{-4} $ \\
					QMT loading & $1.46(7) \times 10^{9} $ & $32(2)$ & $5.6(5) \times 10^{10} $ & $1.3(2) \times 10^{-5} $ \\
					QMT after RF ramp ($2.5 \; \mathrm{s}$) & $0.81(5) \times 10^{9} $ & $32(3)$ & $3.9(4) \times 10^{10} $ & $0.8(2) \times 10^{-5}$ \\
					ODT after RF ramp ($2.5 \; \mathrm{s}$)  & $8.5(2)  \times 10^{6} $  & $40(10)$ & $8.2(8) \times 10^{13} $ & $1.6(9)\times 10^{-2}$\\
					ODT after load & $2.8(3)  \times 10^{7} $  & $39(2)$ & $3.1(4)\times 10^{14} $ & $6(1) \times 10^{-2}$  \\
					%\bottomrule
				\end{tabular}
			\end{ruledtabular}
		\end{center}
	\end{table*}

The hybrid trap configuration is sketched in Fig. \ref{fig:hybridconf}(a). A horizontal single beam ODT is vertically displaced from the QMT trap center, where the magnetic field is zero, by a distance $z_\mathrm{dip}$. As a consequence of this displacement, atoms trapped in the ODT potential avoid depolarization due to Majorana spin flips. The simmetry axis of the quadrupole magnetic potential ($\hat{z}$) is orthogonal to the dipole beam axis ($\hat{x}$). The resulting potential, accounting for gravity, is given by:

\begin{multline} \label{eq:hybridpot}
U(x,y,z)= \mu B'_\mathrm{z} \sqrt{\frac{x^2}{4} + \frac{y^2}{4} + z^2} \\ - C \frac{2P}{ \pi w^2(x)} \exp{ \left \{ -2 \frac{ \left [ y^2 + (z-z_{\mathrm{dip}})^2 \right ]}{w(x)^2} \right \} } + mgz.
\end{multline} The first term in the right-hand side of the equation is the quadrupole magnetic potential, the second is the optical potential term and the third term is the gravitational potential. Here $B'_\mathrm{z}$  is the magnetic field gradient along $\hat{z}$, $\mu$ is the magnetic moment of the atoms, $w(x)$ is the beam radius ($1/e^2$), and $C$ is a constant proportional to the polarizability of the atomic species, depending on the trapping beam wavelength.

The ODT consists of far-detuned ($\lambda_\mathrm{dip}=1064 \; \mathrm{nm}$) laser beam, focused on the atoms through a $f=200 \; \mathrm{mm}$ achromatic lens to a waist $w_0=23 \; \mathrm{ \SIUnitSymbolMicro m}$. The $C$ constant for the given wavelength is $7.27 \, \mathsf{x} \, 10^{-37} \; \mathrm{J/(W \,m^{-2})}$. The ODT is operated at a maximum power of $7.5 \; \mathrm{W}$, corresponding to a trap depth of $U_\mathrm{0}\simeq k_\mathrm{B} \times 475 \; \SIUnitSymbolMicro  \mathrm{K}$, where $k_\mathrm{B}$ is the Boltzmann constant.

Given our beam parameters and magnetic field gradient, the radial trapping frequencies $\omega_\mathrm{y,z}$ are dominated by the ODT contribution 
\begin{equation}
\label{eq:radfreq}
\omega_\mathrm{y,z} \simeq 2 \sqrt{U_\mathrm{0}/m w_0^2},
\end{equation} while both the optical and magnetic potential contribute to the axial trapping frequency
\begin{equation}
\label{eq:axfreq}
\omega_\mathrm{x}=\sqrt{  \mu B'_\mathrm{z} /(4m z_\mathrm{dip}) + \left(\omega^\mathrm{ODT}_\mathrm{x}\right)^2 }. 
\end{equation}

Trap frequencies are experimentally measured by exciting collective modes in the condensate \cite{Stringari96} at the end of the experimental sequence. At the end of the evaporation, we achieve a pure BEC with an ODT depth of $k_\mathrm{B} \times 3.8 \; \SIUnitSymbolMicro \mathrm{K}$, and a QMT gradient of $B'_\mathrm{z}=7.74 \; \mathrm{G/cm}$. Here the radial trapping frequencies are $\{\omega_\mathrm{y},\omega_\mathrm{z}\}=2\pi \times \{512(3),510(3)\} \; \mathrm{Hz}$, while the axial one is reported in Fig. \ref{fig:hybridconf}(b), as a function of the displacement $z_\mathrm{dip}$. The displacement is controlled by applying a homogeneous bias field along the $z$ direction, which results in a vertical shift of the QMT center position $\Delta z_\mathrm{bias}=B_\mathrm{bias,z}/B'_\mathrm{z}$. The axial ODT contribution, extracted by fitting the experimental data with Eq. \eqref{eq:axfreq}, is $\omega^\mathrm{ODT}_\mathrm{x}=2.5(2)\; \mathrm{Hz}$. Such a value is roughly half of the value expected for an ideal Gaussian beam with the given beam parameters, probably as a consequence of non ideal focusing or imperfect beam quality.

\section{\label{sec:expseq}Characterization of the experimental sequence}

\begin{figure}[h!]
	\resizebox{\columnwidth}{!}{
		\includegraphics{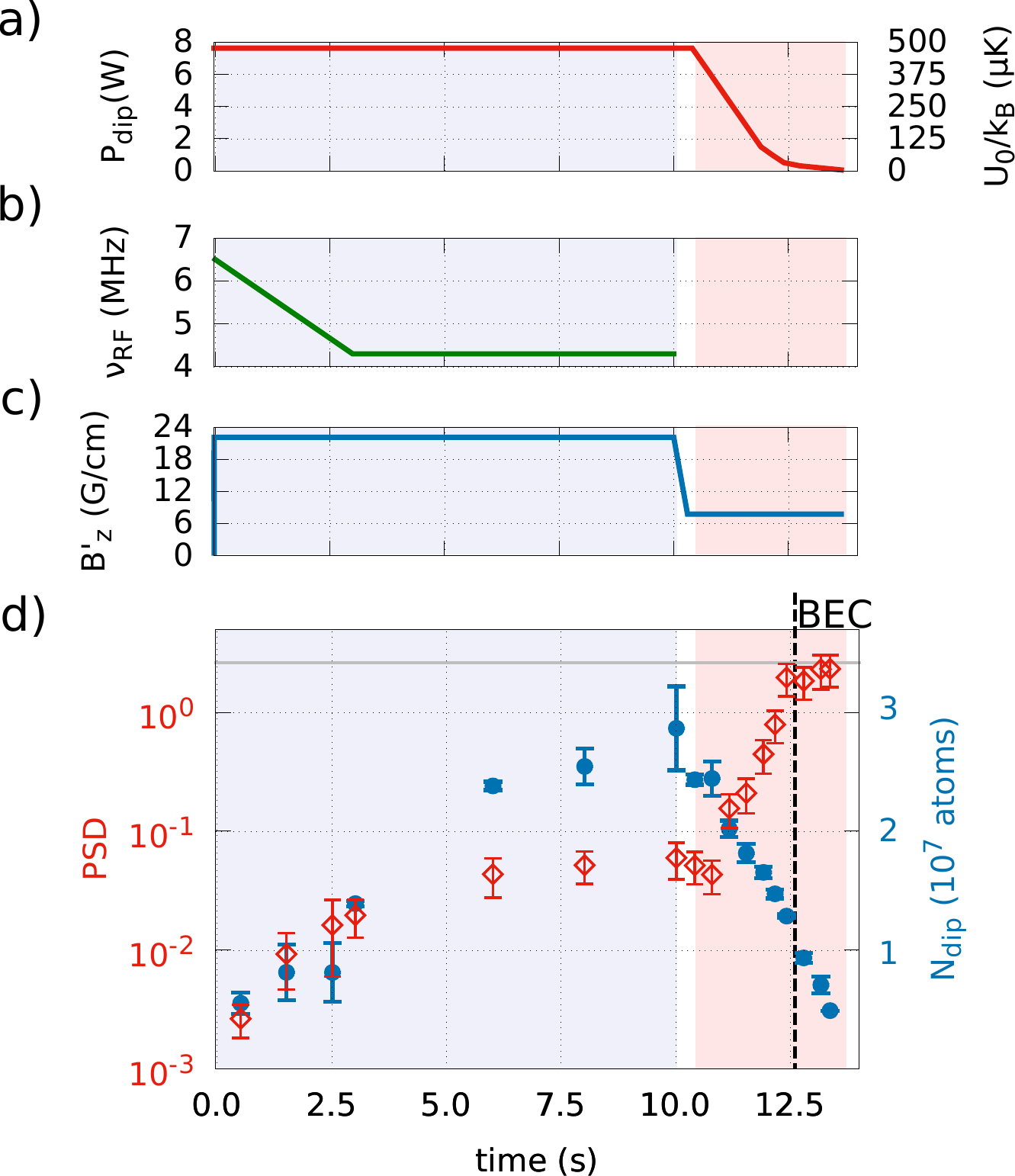}}
	\caption{\label{fig:all} Power of the ODT beam \textbf{(a)}, RF knife frequency \textbf{(b)}, and QMT gradient \textbf{(c)} as a function of time, during the ODT load sequence (gray shaded region) and ODT evaporation (red shaded region). Panel \textbf{(d)} shows the measured atom number (blue filled circles) in ODT, and the estimated PSD value (red empty diamonds). Reported error bars for PSD account for systematic uncertainty in the estimation of trapping frequencies. The gray line signals the PSD value at the condensation threshold PSD = 2.6. Atom numbers reported after BEC threshold account for the total number, while the PSD value accounts for the thermal fraction only.}
\end{figure}

The experimental sequence carried out to produce Bose-Einstein condensates is shown in Fig. \ref{fig:all}(a-c). Right after the end of the GM, the atoms are captured in the combined potential by suddenly switching on the QMT at $22.1 \; \mathrm{G/cm}$ in the presence of the ODT with a trap depth $k_\mathrm{B} \times 475 \; \SIUnitSymbolMicro \mathrm{K}$. About $1.5 \times 10^9$ atoms, corresponding to the fraction occupying the low-field-seeking state $m_\mathrm{F}=-1$, are captured in the quadrupole trap at a temperature of about $32 \; \mathrm{\SIUnitSymbolMicro K}$.

During the subsequent loading stage, lasting  $10 \; \mathrm{s}$, a fraction of the atoms is loaded by elastic collisions into the ODT, where the sample rapidly  becomes collisionally thick in both the radial and axial directions, as the system reaches thermal equilibrium. In the meanwhile, the reservoir temperature is stabilized by applying an RF knife whose frequency is swept from $6.5$ to $4.3 \; \mathrm{MHz}$ in $2.5 \; \mathrm{s}$ and kept constant until the end of the loading stage, where the number of atoms captured in the ODT saturates to a value of the order of $3 \times 10^7$ with a temperature of the order of $ 40 \;\mathrm{\SIUnitSymbolMicro K}$. Table \ref{tab:stages} summarizes the quantitative results up to this point.

After switching off the RF signal, the magnetic field gradient is ramped down in $300 \; \mathrm{ms}$ to a value of $7.74 \; \mathrm{G/cm}$, that is slightly below the one necessary to compensate for gravity, $B'_\mathrm{z} < mg/ \mu \simeq 8.1 \; \mathrm{G/cm}$. In this way the atoms that are held only by the quadrupole potential are released. No significant improvement is observed with longer ramping times. The sample is then evaporatively cooled to quantum degeneracy, with an initial truncation factor $\eta=U_\mathrm{0}/(k_\mathrm{B} T) \simeq 13$, by ramping down the ODT depth from $k_\mathrm{B} \times 475 \;  \SIUnitSymbolMicro \mathrm{K} $ to $k_\mathrm{B} \times 3.8 \; \SIUnitSymbolMicro \mathrm{K}$ in $3.25 \; \mathrm{s}$, after which we obtain an almost pure Bose-Einstein condensate made of about $7$ million atoms. 

Figure \ref{fig:all}(d) shows the number of atoms loaded in the ODT, as well as the corresponding phase-space density ($\mathrm{PSD}$) during the experimental procedure. Such values are estimated assuming thermal equilibrium in trap $\mathrm{PSD}= N (\hbar \bar{\omega})^3/(k_B T)^3$, where $\bar{\omega}=(\omega_\mathrm{x} \omega_\mathrm{y}\omega_\mathrm{z})^{1/3}$ is the instantaneous geometric average of the trapping frequencies. Trapping frequencies at different values of ODT power and magnetic field gradient are estimated from the ones experimentally measured at given conditions, through the dependencies on $U_\mathrm{0}$ and $B'_\mathrm{z}$ in equations \eqref{eq:radfreq} and \eqref{eq:axfreq}.

\begin{figure*}[t!]
	\resizebox{\textwidth}{!}{
		\includegraphics{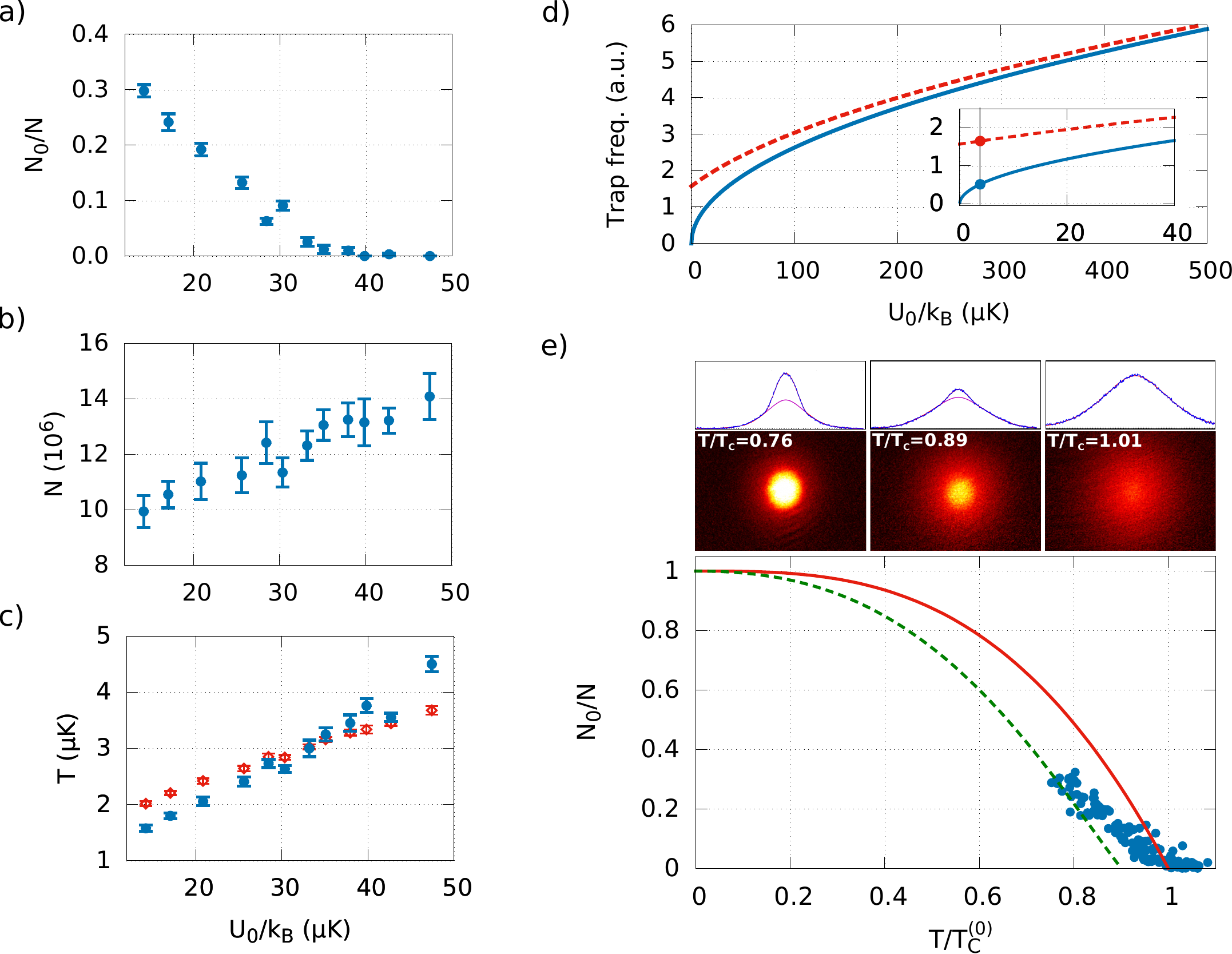}}
	\caption{\label{fig:tcall} Condensed fraction and total atom number as a function of the optical trap depth during evaporation \textbf{(a,b)}. Measured temperature (blue dots) and ideal gas critical temperature $T^\mathrm{(0)}_\mathrm{C}$ (red diamonds) estimated for each data point \textbf{(c)}, using the values of trapping frequencies inferred from the dependence on the ODT power, represented in arbitrary units \textbf{(d)}. Arbitrary frequency units amount to 1 kHz for the radial trapping (blue solid line) and to 5 Hz for the axial trapping (red dashed line). The inset shows the scaling behavior at low values of dipole power, where the effect of the residual magnetic confinement is evident. The gray line signals the final power at the end of the evaporation, where the trapping frequencies are experimentally measured. Panel \textbf{(e)} shows the condensed fraction in the function of the normalized temperature (blue dots) compared to the ideal gas curve (red solid line) and to the curve corrected for mean field interactions \cite{Dalfovo99} (green dashed line). Image profiles for three different temperature values are shown on top.}
\end{figure*}

The system is characterized in the vicinity of the transition temperature by evaporating the sample to different final values of ODT depth $U_\mathrm{0}$, and measuring simultaneously the condensed fraction [Fig. \ref{fig:tcall}(a)], the total atom number [Fig. \ref{fig:tcall}(b)], and the temperature [Fig. \ref{fig:tcall}(c)] of the sample. The harmonically trapped ideal gas transition temperature $T^\mathrm{(0)}_\mathrm{C}=\hbar \bar{\omega} N^{1/3} / [k_\mathrm{B} \zeta^{1/3}(3)]$ is computed for each data point, estimating $\bar{\omega}$ at different ODT powers as explained above, and shown in Fig. \ref{fig:tcall}(d). The BEC phase transition is crossed at an approximate temperature of $3 \; \SIUnitSymbolMicro \mathrm{K}$. In Fig. \ref{fig:tcall}(e), the measured values of condensed fraction are reported as a function of the reduced temperature $T/T^\mathrm{(0)}_\mathrm{C}$. As expected, the experimental points lie below the ideal gas curve, as a consequence of repulsive interactions \cite{Dalfovo99}. The data show general agreement with the curve accounting for mean-field interactions, except in the vicinity of $T^\mathrm{(0)}_\mathrm{C}$ where the approximation is not valid.

\section{\label{sec:conc}Conclusions}
We described a new experimental apparatus for the production of large BECs in a hybrid trap compatible with the use of $\SIUnitSymbolMicro$-metal, or equivalent alloys, magnetic shields. GM cooling allows for an efficient mode-matching with the low-gradient QMT. Atoms captured in the QMT act as a reservoir during the loading of the ODT, which operates in the collisionally thick regime. This procedure allows to efficiently load the atomic sample into the ODT, even in the absence of magnetic trap compression, preparing it in conditions suitable for an efficient optical evaporation, during which the magnetic potential also contributes to thr ODT axial confinement.

The perspectives opened by the present work, together with the following implementation of a properly designed magnetic shield to work in conditions of high magnetic field stability, include the possibility to explore the coherent evolution of Rabi-coupled mixtures subject to first order Zeeman perturbation, on a timescale longer than the orbital many-body dynamics. In particular, the binary mixture composed of the ground hyperfine states $\left| 1,\pm1 \right \rangle$ of $^{23}\mathrm{Na}$ is a promising platform to study the exotic defect structures emerging in Rabi coupled mixtures, such as as magnetic solitons \cite{Qu17} and vortex molecules \cite{Kasamatsu04,Tylutki16,Calderaro17}, as well as supersolid phases exhibited by spin-orbit coupled BECs \cite{Lin11,Li13,Han15,Li17,Leonard17}.

\subsection*{Aknowledgments}
We thank T. Bienaim\'e for useful discussions and E. Iseni for technical assistance in the early stage of the experiment.
This work has been supported by the Provincia Autonoma di Trento, by the QUIC grant of the Horizon 2020 FET program, by INFN, and by NAQUAS project.


\begin{thebibliography}{10}
	
	\bibitem{Bloch12}
	I.Bloch, J.Dalibard, and S.Nascimb\`ene.
	\newblock {Quantum simulations with ultracold quantum gases}.
	\newblock {\em Nature Physics}, 8:267, 2012.
	
	\bibitem{Georgescu14}
	I.~M. Georgescu, S.~Ashhab, and Franco Nori.
	\newblock Quantum simulation.
	\newblock {\em Rev. Mod. Phys.}, 86:153--185, Mar 2014.
	
	\bibitem{Gross17}
	Christian Gross and Immanuel Bloch.
	\newblock Quantum simulations with ultracold atoms in optical lattices.
	\newblock {\em Science}, 357(6355):995--1001, 2017.
	
	\bibitem{Son02}
	D.~T. Son and M.~A. Stephanov.
	\newblock Domain walls of relative phase in two-component {B}ose-{E}instein
	condensates.
	\newblock {\em Phys. Rev. A}, 65:063621, 2002.
	
	\bibitem{Kasamatsu04}
	Kenichi Kasamatsu, Makoto Tsubota, and Masahito Ueda.
	\newblock Vortex molecules in coherently coupled two-component
	{B}ose-{E}instein condensates.
	\newblock {\em Phys. Rev. Lett.}, 93:250406, 2004.
	
	\bibitem{Kasamatsu05}
	Kenichi Kasamatsu, Makoto Tsubota, and Masahito Ueda.
	\newblock Vortices in multicomponent {B}ose-{E}instein condensates.
	\newblock {\em International Journal of Modern Physics B}, 19(11):1835--1904,
	2005.
	
	\bibitem{Cipriani13}
	Mattia Cipriani and Muneto Nitta.
	\newblock Crossover between integer and fractional vortex lattices in
	coherently coupled two-component {B}ose-{E}instein condensates.
	\newblock {\em Phys. Rev. Lett.}, 111:170401, 2013.
	
	\bibitem{Tylutki16}
	Marek Tylutki, Lev~P. Pitaevskii, Alessio Recati, and Sandro Stringari.
	\newblock Confinement and precession of vortex pairs in coherently coupled
	{B}ose-{E}instein condensates.
	\newblock {\em Phys. Rev. A}, 93:043623, Apr 2016.
	
	\bibitem{Aftalion16}
	Amandine Aftalion and Peter Mason.
	\newblock Rabi-coupled two-component {B}ose-{E}instein condensates:
	Classification of the ground states, defects, and energy estimates.
	\newblock {\em Phys. Rev. A}, 94:023616, 2016.
	
	\bibitem{Qu17}
	Chunlei Qu, Marek Tylutki, Sandro Stringari, and Lev~P. Pitaevskii.
	\newblock Magnetic solitons in {R}abi-coupled {B}ose-{E}instein condensates.
	\newblock {\em Phys. Rev. A}, 95:033614, 2017.
	
	\bibitem{Calderaro17}
	Luca Calderaro, Alexander~L. Fetter, Pietro Massignan, and Peter Wittek.
	\newblock Vortex dynamics in coherently coupled {B}ose-{E}instein condensates.
	\newblock {\em Phys. Rev. A}, 95:023605, 2017.
	
	\bibitem{Eto17}
	Minoru Eto and Muneto Nitta.
	\newblock Confinement of half-quantized vortices in coherently coupled
	{B}ose-{E}instein condensates: Simulating quark confinement in a {QCD}-like
	theory.
	\newblock {\em Phys. Rev. A}, 97:023613, 2018.
	
	\bibitem{Juzeliunas10}
	Gediminas Juzeli\ifmmode~\bar{u}\else \={u}\fi{}nas, Julius Ruseckas, and Jean
	Dalibard.
	\newblock Generalized {R}ashba-{D}resselhaus spin-orbit coupling for cold
	atoms, 2010.
	
	\bibitem{Brandon10}
	Brandon~M. Anderson, Gediminas Juzeli\ifmmode~\bar{u}\else \={u}\fi{}nas,
	Victor~M. Galitski, and I.~B. Spielman.
	\newblock Synthetic 3{D} spin-orbit coupling.
	\newblock {\em Phys. Rev. Lett.}, 108:235301, Jun 2012.
	
	\bibitem{Li12}
	Yun Li, Lev~P. Pitaevskii, and Sandro Stringari.
	\newblock Quantum tricriticality and phase transitions in spin-orbit coupled
	{B}ose-{E}instein condensates.
	\newblock {\em Phys. Rev. Lett.}, 108:225301, May 2012.
	
	\bibitem{Li12b}
	Yun Li, Giovanni~Italo Martone, and Sandro Stringari.
	\newblock Sum rules, dipole oscillation and spin polarizability of a spin-orbit
	coupled quantum gas.
	\newblock {\em EPL (Europhysics Letters)}, 99(5):56008, 2012.
	
	\bibitem{Li13}
	Yun Li, Giovanni~I. Martone, Lev~P. Pitaevskii, and Sandro Stringari.
	\newblock Superstripes and the excitation spectrum of a spin-orbit-coupled
	{B}ose-{E}instein condensate.
	\newblock {\em Phys. Rev. Lett.}, 110:235302, Jun 2013.
	
	\bibitem{Han15}
	Wei Han, Gediminas Juzeli\ifmmode~\bar{u}\else \={u}\fi{}nas, Wei Zhang, and
	Wu-Ming Liu.
	\newblock Supersolid with nontrivial topological spin textures in
	spin-orbit-coupled {B}ose gases.
	\newblock {\em Phys. Rev. A}, 91:013607, Jan 2015.
	
	\bibitem{Li17}
	Jun-Ru Li, Jeongwon Lee, Wujie Huang, Sean Burchesky, Boris Shteynas,
	Furkan~{\c{C}}a{\u{g}}r{\i} Top, Alan~O Jamison, and Wolfgang Ketterle.
	\newblock A stripe phase with supersolid properties in spin--orbit-coupled
	{B}ose--{E}instein condensates.
	\newblock {\em Nature}, 543(7643):91, 2017.
	
	\bibitem{Leonard17}
	Julian L{\'e}onard, Andrea Morales, Philip Zupancic, Tilman Esslinger, and
	Tobias Donner.
	\newblock Supersolid formation in a quantum gas breaking a continuous
	translational symmetry.
	\newblock {\em Nature}, 543(7643):87, 2017.
	
	\bibitem{Butera17}
	Salvatore Butera, Patrik \"Ohberg, and Iacopo Carusotto.
	\newblock Black-hole lasing in coherently coupled two-component atomic
	condensates.
	\newblock {\em Phys. Rev. A}, 96:013611, 2017.
	
	\bibitem{Harber02}
	D.~M. Harber, H.~J. Lewandowski, J.~M. McGuirk, and E.~A. Cornell.
	\newblock Effect of cold collisions on spin coherence and resonance shifts in a
	magnetically trapped ultracold gas.
	\newblock {\em Phys. Rev. A}, 66:053616, 2002.
	
	\bibitem{Treutlein04}
	Philipp Treutlein, Peter Hommelhoff, Tilo Steinmetz, Theodor~W. H\"ansch, and
	Jakob Reichel.
	\newblock Coherence in microchip traps.
	\newblock {\em Phys. Rev. Lett.}, 92:203005, 2004.
	
	\bibitem{Deutsch10}
	C.~Deutsch, F.~Ramirez-Martinez, C.~Lacro\^ute, F.~Reinhard, T.~Schneider,
	J.~N. Fuchs, F.~Pi\'echon, F.~Lalo\"e, J.~Reichel, and P.~Rosenbusch.
	\newblock Spin self-rephasing and very long coherence times in a trapped atomic
	ensemble.
	\newblock {\em Phys. Rev. Lett.}, 105:020401, 2010.
	
	\bibitem{Kleine11}
	G.~Kleine~B\"uning, J.~Will, W.~Ertmer, E.~Rasel, J.~Arlt, C.~Klempt,
	F.~Ramirez-Martinez, F.~Pi\'echon, and P.~Rosenbusch.
	\newblock Extended coherence time on the clock transition of optically trapped
	rubidium.
	\newblock {\em Phys. Rev. Lett.}, 106:240801, 2011.
	
	\bibitem{Muessel15}
	W.~Muessel, H.~Strobel, D.~Linnemann, T.~Zibold, B.~Juli\'a-D\'{\i}az, and
	M.~K. Oberthaler.
	\newblock Twist-and-turn spin squeezing in bose-einstein condensates.
	\newblock {\em Phys. Rev. A}, 92:023603, Aug 2015.
	
	\bibitem{Bienaime16}
	Tom Bienaim\'e, Eleonora Fava, Giacomo Colzi, Carmelo Mordini, Simone Serafini,
	Chunlei Qu, Sandro Stringari, Giacomo Lamporesi, and Gabriele Ferrari.
	\newblock Spin-dipole oscillation and polarizability of a binary
	{B}ose-{E}instein condensate near the miscible-immiscible phase transition.
	\newblock {\em Phys. Rev. A}, 94:063652, 2016.
	
	\bibitem{Fava17}
	E.~{Fava}, T.~{Bienaim{\'e}}, C.~{Mordini}, G.~{Colzi}, C.~{Qu},
	S.~{Stringari}, G.~{Lamporesi}, and G.~{Ferrari}.
	\newblock {Spin Superfluidity of a Bose Gas Mixture at Finite Temperature}.
	\newblock arXiv:1708.03923 [cond-mat.quant-gas], 2017.
	
	\bibitem{Krivanek08}
	O.L. Krivanek, G.J. Corbin, N.~Dellby, B.F. Elston, R.J. Keyse, M.F. Murfitt,
	C.S. Own, Z.S. Szilagyi, and J.W. Woodruff.
	\newblock An electron microscope for the aberration-corrected era.
	\newblock {\em Ultramicroscopy}, 108(3):179 -- 195, 2008.
	
	\bibitem{Mansfield87}
	P~Mansfield and B~Chapman.
	\newblock Multishield active magnetic screening of coil structures in {NMR}.
	\newblock {\em Journal of Magnetic Resonance (1969)}, 72(2):211 -- 223, 1987.
	
	\bibitem{Ottl06}
	Anton \"Ottl, Stephan Ritter, Michael K\"ohl, and Tilman Esslinger.
	\newblock Hybrid apparatus for {B}ose-{E}instein condensation and cavity
	quantum electrodynamics: Single atom detection in quantum degenerate gases.
	\newblock {\em Review of Scientific Instruments}, 77(6):063118, 2006.
	
	\bibitem{Dedman07}
	C.~J. Dedman, R.~G. Dall, L.~J. Byron, and A.~G. Truscott.
	\newblock Active cancellation of stray magnetic fields in a {B}ose-{E}instein
	condensation experiment.
	\newblock {\em Review of Scientific Instruments}, 78(2):024703, 2007.
	
	\bibitem{Zhang15}
	Wenxian Zhang, S.~Yi, M.~S. Chapman, and J.~Q. You.
	\newblock Coherent zero-field magnetization resonance in a dipolar spin-1
	{B}ose-{E}instein condensate.
	\newblock {\em Phys. Rev. A}, 92:023615, 2015.
	
	\bibitem{Sheng13}
	D.~Sheng, S.~Li, N.~Dural, and M.~V. Romalis.
	\newblock Subfemtotesla scalar atomic magnetometry using multipass cells.
	\newblock {\em Phys. Rev. Lett.}, 110:160802, 2013.
	
	\bibitem{VanZoest10}
	T.~van Zoest, N.~Gaaloul, Y.~Singh, H.~Ahlers, W.~Herr, S.~T. Seidel,
	W.~Ertmer, E.~Rasel, M.~Eckart, E.~Kajari, S.~Arnold, G.~Nandi, W.~P.
	Schleich, R.~Walser, A.~Vogel, K.~Sengstock, K.~Bongs, W.~Lewoczko-Adamczyk,
	M.~Schiemangk, T.~Schuldt, A.~Peters, T.~K{\"o}nemann, H.~M{\"u}ntinga,
	C.~L{\"a}mmerzahl, H.~Dittus, T.~Steinmetz, T.~W. H{\"a}nsch, and J.~Reichel.
	\newblock {B}ose-{E}instein condensation in microgravity.
	\newblock {\em Science}, 328(5985):1540--1543, 2010.
	
	\bibitem{Milke14}
	Alexander Milke, Andr\'e Kubelka-Lange, Norman G\"urlebeck, Benny Rievers, Sven
	Herrmann, Thilo Schuldt, and Claus Braxmaier.
	\newblock Atom interferometry in space: Thermal management and magnetic
	shielding.
	\newblock {\em Review of Scientific Instruments}, 85(8):083105, 2014.
	
	\bibitem{Kubelka-Lange16}
	Andr\'e Kubelka-Lange, Sven Herrmann, Jens Grosse, Claus Lämmerzahl, Ernst~M.
	Rasel, and Claus Braxmaier.
	\newblock A three-layer magnetic shielding for the {MAIUS}-1 mission on a
	sounding rocket.
	\newblock {\em Review of Scientific Instruments}, 87(6):063101, 2016.
	
	\bibitem{deAngelis09}
	M~de~Angelis, A~Bertoldi, L~Cacciapuoti, A~Giorgini, G~Lamporesi, M~Prevedelli,
	G~Saccorotti, F~Sorrentino, and G~M Tino.
	\newblock Precision gravimetry with atomic sensors.
	\newblock {\em Measurement Science and Technology}, 20(2):022001, 2009.
	
	\bibitem{Lamporesi08}
	G.~Lamporesi, A.~Bertoldi, L.~Cacciapuoti, M.~Prevedelli, and G.~M. Tino.
	\newblock Determination of the newtonian gravitational constant using atom
	interferometry.
	\newblock {\em Phys. Rev. Lett.}, 100:050801, 2008.
	
	\bibitem{Dickerson13}
	Susannah~M. Dickerson, Jason~M. Hogan, Alex Sugarbaker, David M.~S. Johnson,
	and Mark~A. Kasevich.
	\newblock Multiaxis inertial sensing with long-time point source atom
	interferometry.
	\newblock {\em Phys. Rev. Lett.}, 111:083001, 2013.
	
	\bibitem{Hartwig15}
	J~Hartwig, S~Abend, C~Schubert, D~Schlippert, H~Ahlers, K~Posso-Trujillo,
	N~Gaaloul, W~Ertmer, and E~M Rasel.
	\newblock Testing the universality of free fall with rubidium and ytterbium in
	a very large baseline atom interferometer.
	\newblock {\em New Journal of Physics}, 17(3):035011, 2015.
	
	\bibitem{Botti06}
	Laura Botti, Roberto Buffa, Andrea Bertoldi, Davide Bassi, and Leonardo Ricci.
	\newblock Noninvasive system for the simultaneous stabilization and control of
	magnetic field strength and gradient.
	\newblock {\em Review of Scientific Instruments}, 77(3):035103, 2006.
	
	\bibitem{shieldprep}
	\rm G. Colzi~et al.
	\newblock in preparation.
	
	\bibitem{Esslinger98}
	Tilman Esslinger, Immanuel Bloch, and Theodor~W. H\"ansch.
	\newblock {B}ose-{E}instein condensation in a quadrupole-{I}offe-configuration
	trap.
	\newblock {\em Phys. Rev. A}, 58:R2664--R2667, 1998.
	
	\bibitem{Bloch99}
	Immanuel Bloch, Theodor~W. H\"ansch, and Tilman Esslinger.
	\newblock Atom laser with a cw output coupler.
	\newblock {\em Phys. Rev. Lett.}, 82:3008--3011, 1999.
	
	\bibitem{blochcom}
	\rm I.~Bloch.
	\newblock Private communication.
	
	\bibitem{Kim13}
	Huidong Kim, Hyok~Sang Han, and D.~Cho.
	\newblock Magic polarization for optical trapping of atoms without
	{S}tark-induced dephasing.
	\newblock {\em Phys. Rev. Lett.}, 111:243004, Dec 2013.
	
	\bibitem{Sycz18}
	Krystian Sycz, Adam~M Wojciechowski, and Wojciech Gawlik.
	\newblock Atomic-state diagnostics and optimization in cold-atom experiments.
	\newblock {\em Scientific reports}, 8(1):2805, 2018.
	
	\bibitem{Davis95}
	K.~B. Davis, M.~O. Mewes, M.~R. Andrews, N.~J. van Druten, D.~S. Durfee, D.~M.
	Kurn, and W.~Ketterle.
	\newblock {B}ose-{E}instein condensation in a gas of sodium atoms.
	\newblock {\em Phys. Rev. Lett.}, 75:3969--3973, 1995.
	
	\bibitem{Naik05}
	D.~S. Naik and C.~Raman.
	\newblock Optically plugged quadrupole trap for {B}ose-{E}instein condensates.
	\newblock {\em Phys. Rev. A}, 71:033617, 2005.
	
	\bibitem{Heo11}
	Myoung-Sun Heo, Jae-yoon Choi, and Yong-il Shin.
	\newblock Fast production of large $^{23}\mathrm{Na}$ {B}ose-{E}instein
	condensates in an optically plugged magnetic quadrupole trap.
	\newblock {\em Phys. Rev. A}, 83:013622, 2011.
	
	\bibitem{Dubessy12}
	R.~Dubessy, K.~Merloti, L.~Longchambon, P.-E. Pottie, T.~Liennard, A.~Perrin,
	V.~Lorent, and H.~Perrin.
	\newblock Rubidium-87 {B}ose-{E}instein condensate in an optically plugged
	quadrupole trap.
	\newblock {\em Phys. Rev. A}, 85:013643, 2012.
	
	\bibitem{Lin09}
	Y.-J. Lin, A.~R. Perry, R.~L. Compton, I.~B. Spielman, and J.~V. Porto.
	\newblock Rapid production of $^{87}\text{R}\text{b}$ {B}ose-{E}instein
	condensates in a combined magnetic and optical potential.
	\newblock {\em Phys. Rev. A}, 79:063631, 2009.
	
	\bibitem{Flores15}
	Adonis~Silva Flores, Hari~Prasad Mishra, Wim Vassen, and Steven Knoop.
	\newblock Simple method for producing {B}ose--{E}instein condensates of
	metastable helium using a single-beam optical dipole trap.
	\newblock {\em Applied Physics B}, 121(3):391--399, 2015.
	
	\bibitem{Mishra15}
	Hari~Prasad Mishra, Adonis~Silva Flores, Wim Vassen, and Steven Knoop.
	\newblock Efficient production of an $^{87}${R}b {F} = 2, m$_{F}$ = 2
	{B}ose-{E}instein condensate in a hybrid trap.
	\newblock {\em The European Physical Journal D}, 69(2):52, 2015.
	
	\bibitem{Bouton15}
	Q.~Bouton, R.~Chang, A.~L. Hoendervanger, F.~Nogrette, A.~Aspect, C.~I.
	Westbrook, and D.~Cl\'ement.
	\newblock Fast production of {B}ose-{E}instein condensates of metastable
	helium.
	\newblock {\em Phys. Rev. A}, 91:061402, 2015.
	
	\bibitem{Zaiser11}
	M.~Zaiser, J.~Hartwig, D.~Schlippert, U.~Velte, N.~Winter, V.~Lebedev,
	W.~Ertmer, and E.~M. Rasel.
	\newblock Simple method for generating {B}ose-{E}instein condensates in a weak
	hybrid trap.
	\newblock {\em Phys. Rev. A}, 83:035601, Mar 2011.
	
	\bibitem{Salomon13}
	G.~Salomon, L.~Fouch\'e, P.~Wang, A.~Aspect, P.~Bouyer, and T.~Bourdel.
	\newblock Gray-molasses cooling of $^{39}$ {K} to a high phase-space density.
	\newblock {\em EPL (Europhysics Letters)}, 104(6):63002, 2013.
	
	\bibitem{Kinoshita05}
	Toshiya Kinoshita, Trevor Wenger, and David~S. Weiss.
	\newblock All-optical {B}ose-{E}instein condensation using a compressible
	crossed dipole trap.
	\newblock {\em Phys. Rev. A}, 71:011602, 2005.
	
	\bibitem{Lamporesi13}
	G~Lamporesi, S~Donadello, S~Serafini, and G~Ferrari.
	\newblock Compact high-flux source of cold sodium atoms.
	\newblock {\em Rev. Sci. Instrum.}, 84(6):063102, 2013.
	
	\bibitem{Colzi16}
	G.~Colzi, G.~Durastante, E.~Fava, S.~Serafini, G.~Lamporesi, and G.~Ferrari.
	\newblock Sub-doppler cooling of sodium atoms in gray molasses.
	\newblock {\em Phys. Rev. A}, 93(023421), 2016.
	
	\bibitem{Ketterle93}
	W.~Ketterle, K.B. Davis, M.A. Joffe, A.~Martin, and D.E. Pritchard.
	\newblock High densities of cold atoms in a dark spontaneous-force optical
	trap.
	\newblock {\em Physical review letters}, 70(15):2253--2256, 1993.
	
	\bibitem{Stringari96}
	S.~Stringari.
	\newblock Collective excitations of a trapped {B}ose-condensed gas.
	\newblock {\em Phys. Rev. Lett.}, 77:2360--2363, Sep 1996.
	
	\bibitem{Dalfovo99}
	Franco Dalfovo, Stefano Giorgini, Lev~P. Pitaevskii, and Sandro Stringari.
	\newblock Theory of {B}ose-{E}instein condensation in trapped gases.
	\newblock {\em Rev. Mod. Phys.}, 71:463--512, 1999.
	
	\bibitem{Lin11}
	Y-J Lin, K~Jim{\'e}nez-Garc{\'\i}a, and Ian~B Spielman.
	\newblock Spin--orbit-coupled {B}ose--{E}instein condensates.
	\newblock {\em Nature}, 471(7336):83, 2011.
	
\end{thebibliography}
\end{document}